\documentclass[twocolumn,showpacs,preprintnumbers,amsmath,amssymb,superscriptaddress,prb]{revtex4}
\usepackage{graphicx}
\usepackage{bm}
\begin{document}
\title{Transport Properties in Ferromagnetic Josephson Junction between Triplet Superconductors}
\author{Yousef Rahnavard}
\address{Department of Physics, Faculty of Sciences, University of Isfahan, 81744 Isfahan, Iran}
\author{Gholamreza Rashedi}
\address{Department of Physics, Faculty of Sciences, University of Isfahan, 81744 Isfahan, Iran}
\author{Takehito Yokoyama}
\address{Department of Physics, Tokyo Institute of Technology, 2-12-1 Ookayama, Meguro-ku, Tokyo 152-8551, Japan}
\date{\today}
\begin{abstract}
Charge and spin Josephson currents in a ballistic superconductor-ferromagnet-superconductor
junction with spin-triplet pairing symmetry are studied using the quasiclassical Eilenberger equation.
The gap vector of superconductors has an arbitrary relative angle with respect to magnetization
of the ferromagnetic layer. We clarify the effects of the thickness of ferromagnetic layer and magnitude of the magnetization on the Josephson charge and spin currents.
We find that 0-$\pi$ transition can occur except for the case that the exchange field and d-vector  are in nearly perpendicular configuration.
We also show how spin current flows due to misorientation between the exchange field and d-vector.

\end{abstract}
\pacs{74.50.+r, 74.70.Pq, 74.70.Tx, 72.25.-b} \maketitle

\section{Introduction}
\label{SEC1}
Since both of superconductivity and ferromagnetism are antagonistic ordered phases of matter, for a few decades
after discovery of superconductivity, the interplay between superconductivity and ferromagnetism had not been a
subject of intensive research interest. However, recently the study of interplay between superconductor and ferromagnet
has been revived and, in particular, proximity effect and Andreev reflection in superconductor-ferromagnet junctions has
attracted much attention.\cite{buzdinrmp,Bergeret1,Volkov,Zareyan1,Braude}

When a singlet superconductor has placed in proximity to a
ferromagnetic layer, one can find triplet pairing correlations in the
ferromagnetic layer and this component of pairing provides interesting
effects.\cite{Efetov2,Kadigro,EschrigLTP,eschrig,Yokoyama2,Linder,Alidoust}
On the other hand, spin-triplet superconductor have been extensively
investigated due to its anomalous
features.\cite{Maeno,Ishida,Luke,Mackenzie,
Nelson,Asano1,Tou,Muller,Qian,Abrikosov,Fukuyama,Lebed,Saxena,Pfleiderer,Aoki,Bolech,
Gronsleth,Yasuhiro,Rashedi1,Rashedi2,Rahnavard,Mahmoodi,Kolesnichenko1}
Proximity of spin-triplet superconductor and ferromagnet is more
interesting than the singlet one in the sense that both of
spin-triplet superconductor and ferromagnet have a magnetic nature.

\begin{figure}[tbp]
\includegraphics[width=1.1\columnwidth]{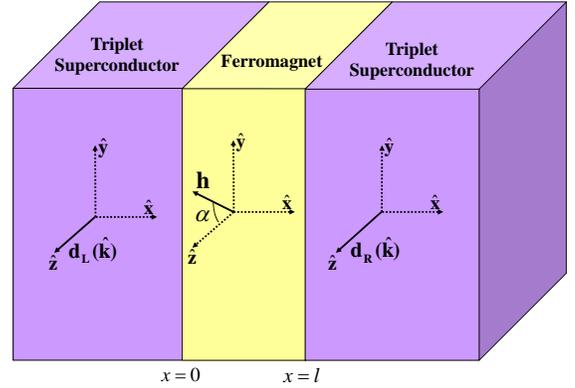}
\caption{(Color online) Scheme of triplet SFS junction. Here, L,R label the left and right
half-spaces. Two superconducting bulks are separated by a ferromagnetic layer with the exchange field $\mathbf{h}$.} \label{fig1}
\end{figure}

In Refs.\cite{Kastening,Brydon1,Brydon2} the authors demonstrated presence
of both charge and spin currents in the systems consisting of  a thin barrier
of ferromagnet sandwiched between two triplet superconductors.
They obtained a spin current due to the coupling of the ferromagnetic moment
to the spin of triplet Cooper pairs.
Also, they found a 0-$\pi$ transition in the triplet
superconductor-ferromagnet-superconductor (SFS) Josephson junction generated by the misalignment of the
$d$-vectors of the triplet superconductors with respect to the magnetic moment of the ferromagnet.
In contrast to Refs.\cite{Kastening,Brydon1,Brydon2} where the ferromagnet
is modeled as $\delta$ function potential,
in this paper we allow for arbitrary length of the ferromagnet,
which is a more realistic modeling, based on the
quasiclassical Eilenberger equation.

In this paper, we investigate a triplet SFS structure with arbitrary misalignment between magnetization of the
ferromagnet and gap vector of superconductors (Fig.\ref{fig1}), by changing the exchange field and thickness of ferromagnetic layer.
We find that charge and spin currents show 0-$\pi$ transition except for the case that the two vectors are in nearly perpendicular configuration

The organization of this paper is as follows. In Sec.\ref{2} the quasiclassical equations for Green functions are presented.
In Sec.\ref{3}, we calculate Green functions of the system analytically, presenting the formulas for the Green functions to
calculate the charge and spin current densities. In Sec.\ref{4}, we show the results of spin and charge currents.
The paper will be finished with the conclusions in Sec.\ref{5}.

\begin{figure}[tbp]
\includegraphics[width=1.1\columnwidth]{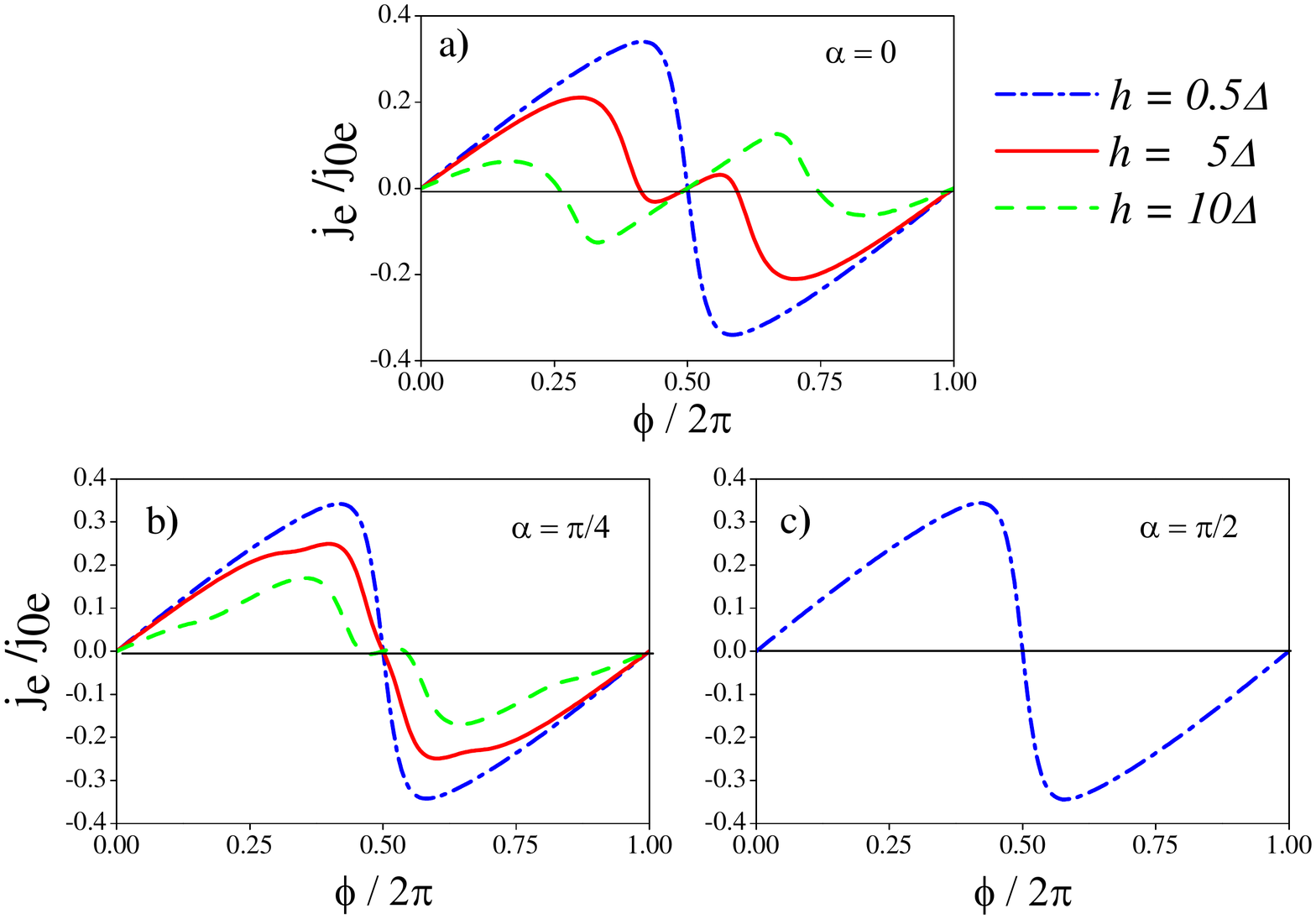}
\caption{(Color online)The $x$ component of the charge
current  versus the phase difference $ \protect\phi $
for $l=\xi/2$, $T=0.05T_{C}$, and
different $h$ and  $\alpha$. Currents are given in units of
$j_{0,e}=\frac{\protect\pi }{2}eN(0)v_{F}\Delta(0) $.}
\label{fig2}
\end{figure}

\begin{figure}[tbp]
\includegraphics[width=1.1\columnwidth]{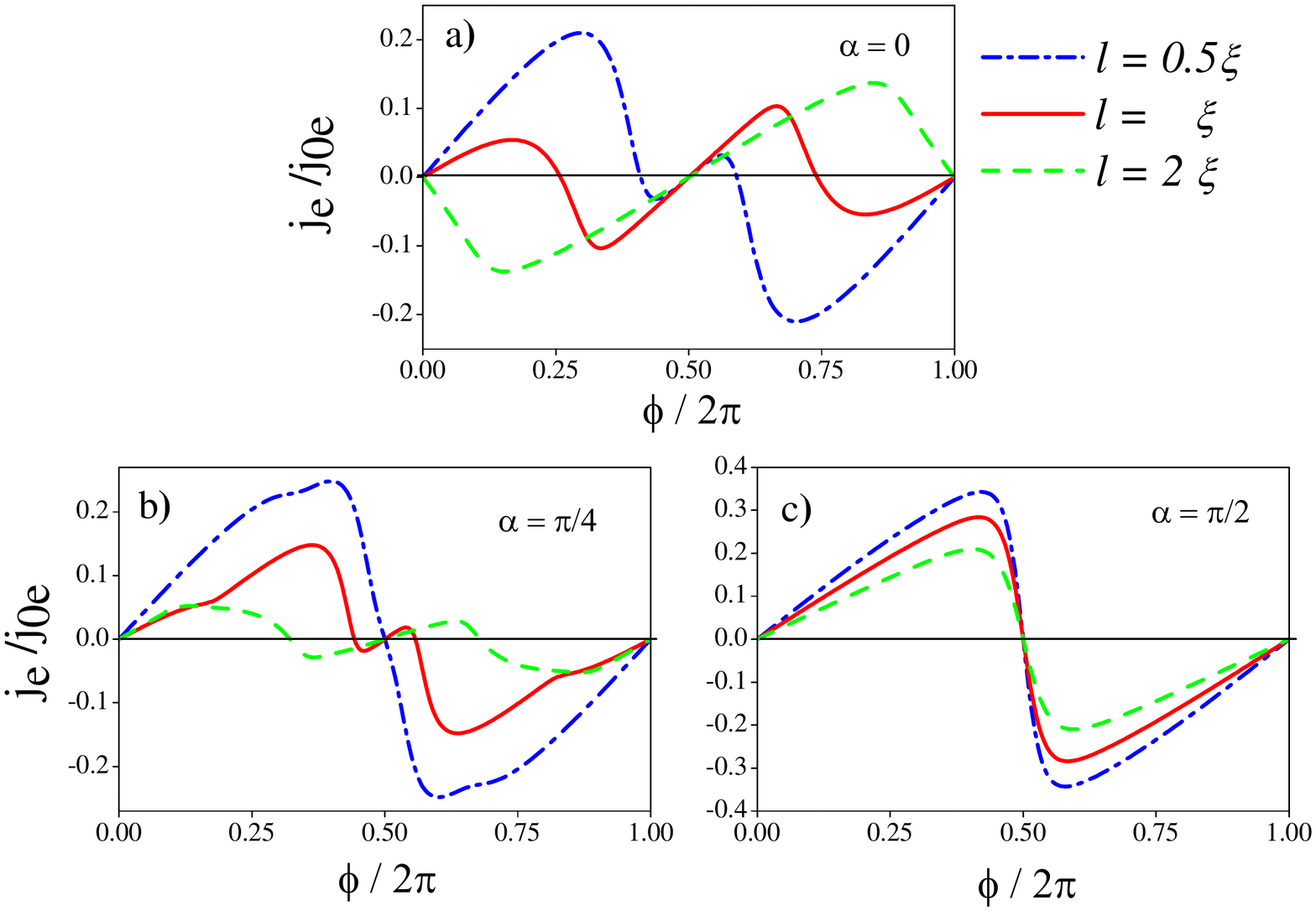}
\caption{(Color online) The $x$ component of the charge
current  versus the phase difference $ \protect\phi $
for $h=5 \Delta$, $T=0.05T_{C}$, and
different $l$ and  $\alpha$.}
\label{fig3}
\end{figure}

\section{Formalism and Basic equations}\label{2}
We consider a clean SFS structure with a homogenous ferromagnet
of thickness $l$ and bulk triplet superconductors(see
Fig.\ref{fig1}). The ferromagnetic layer is characterized by an
exchange field $\mathbf{h}$. There is a misorientation
angle $ \alpha$ between the exchange field of ferromagnet and
order parameters of superconductors ($\mathbf{d}$-vectors).
Interfaces are fully transparent and magnetically inactive. The
thickness $l$ is larger than the Fermi wave length and smaller
than the elastic mean free path. Then, we can use a quasiclassical
description in the clean limit, and apply the Eilenberger equation
in this limit\cite{Eilenberger,Kulik,Kupriyanov}:
\begin{equation}
\mathbf{v}_{F}\nabla \breve{g}+\left[
\varepsilon_{m}\breve{\sigma}_{3}+i(\breve{\Delta}+\breve{\mathbf{h}}),\breve{g}\right] =0, \label{Eilenberger}
\end{equation}
with the normalization condition $\breve{g}\breve{g}=\breve{I}$.
Here, $\varepsilon _{m}=\pi T(2m+1)$ are discrete Matsubara energies, 
$T$ is the temperature, $\mathbf{v}_{F}$ is the
Fermi velocity and $\breve{\sigma}_{3}=\hat{\tau}_{3}\otimes \hat{I}$ in which
$\hat{\tau}_3$ is the third Pauli matrix in particle-hole space.
$\hat{\sigma}_{j}\left(j=1,2,3\right)$  denote Pauli matrices in
spin space in the following.
The Matsubara propagator $\breve{g}$ can be written in the
standard form:
\begin{equation}
\breve{g}=\left(
\begin{array}{cc}
g_{1}+\mathbf{g}_{1}\cdot\hat{\bm{\sigma}} & \left( g_{2}+\mathbf{g}_{2}\cdot\hat{%
\bm{\sigma }}\right) i\hat{\sigma}_{2} \\
i\hat{\sigma}_{2}\left( g_{3}+\mathbf{g}_{3}\cdot\hat{\bm{\sigma
}}\right)  &
g_{4}-\hat{\sigma}_{2}(\mathbf{g}_{4}\cdot\hat{\bm{\sigma
}})\hat{\sigma}_{2}
\end{array}
\right)
\end{equation}\label{Green's function}
where $\hat{\bm\sigma}=(\hat{\sigma}_1,\hat{\sigma}_2,\hat{\sigma}_3)$ is the vector of Pauli matrices in spin space. Also, we can write $\breve{\mathbf{h}}$ as follows:
\begin{equation}
\breve{\mathbf{h}}=\left(
\begin{array}{cc}
\mathbf{h}\cdot\hat{\bm{\sigma}} & 0 \\
0  &
(\mathbf{h}\cdot\hat{\bm{\sigma}})^*
\end{array}
\right).
\end{equation}
\label{ferro matrix}

 We use Eilenberger equation for both ferromagnetic and superconducting materials.
 For superconductors we set $\mathbf{h}=0$.
 The matrix structure of the off-diagonal self energy $\breve{%
\Delta}$ in the Nambu space is
\begin{equation}
\breve{\Delta}=\left(
\begin{array}{cc}
0 & (\mathbf{d}\cdot\hat{\bm{\sigma }})i\hat{\sigma}_{2} \\
i\hat{\sigma}_{2}(\mathbf{{d^{\ast }}\cdot\hat{\bm{\sigma}})} & 0
\end{array}
\right) .
\end{equation}
\label{order parameter}
In the ferromagnet we set $\breve{\Delta}=0$ and $\mathbf{h}=(0,h \sin\alpha,h\cos\alpha)$.
In this paper, the unitary state, $\mathbf{%
d\times d}^{\ast }=0,$ is investigated.

Solutions of Eq. (\ref{Eilenberger})
have to satisfy the conditions for Green functions in the bulks of the superconductors:
\begin{eqnarray}
\breve{g}\left( \pm \infty \right)=\frac{\varepsilon _{m}\breve{\sigma}%
_{3}+i\breve{\Delta}_{1,2}}{\sqrt{\varepsilon _{m}^{2}+\left| \mathbf{d}%
_{1,2}\right| ^{2}}},  \label{Bulk solution} \\
 \mathbf{d}_{1,2}=\mathbf{d}_{L,R}\left( \mathbf{\hat{v%
}}_{F}\right) \exp \left( \pm \frac{i\phi }{2}\right), \label{Bulk
order parameter}
\end{eqnarray}
where $\phi $ is the external phase difference between the order
parameters of the bulks of superconductors. Eq. (\ref{Eilenberger})
have to be supplemented by the continuity conditions at the interfaces
between ferromagnet and superconductors. For all quasiparticle
trajectories, the Green functions satisfy the boundary conditions
both in the right and left bulks as well as at the interfaces.

\begin{figure}[tbp]
\includegraphics[width=1.1\columnwidth]{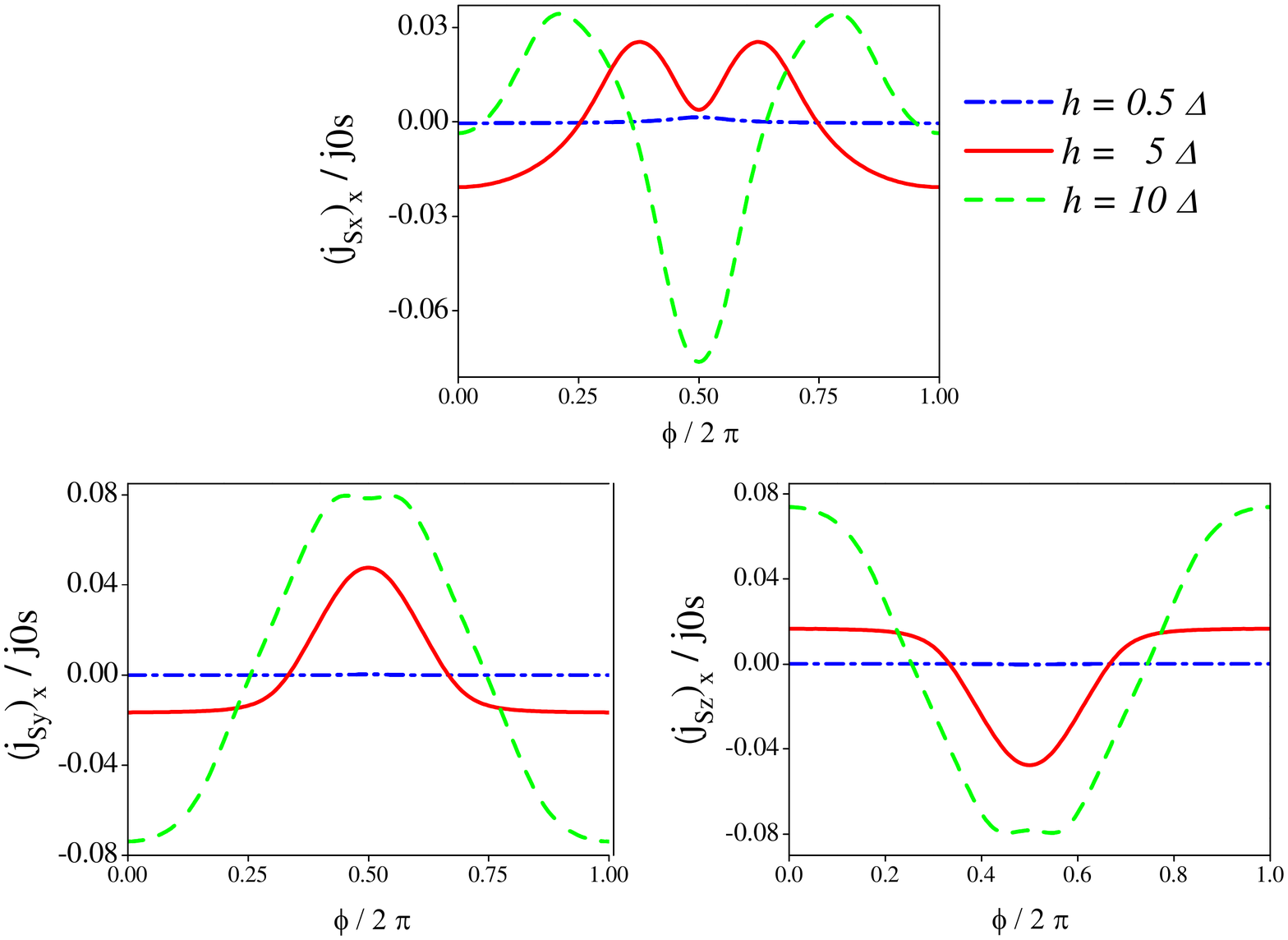}
\caption{(Color online) The normal component ($x$ component) of
the spin current at the left interface, versus the phase difference $
\protect\phi $, for $l=\xi/2$, $T=0.05T_{C}$, different $h$ and $\alpha=\pi/4$.
 Currents are calculated in units of $j_{0,s}=\frac{\protect\pi}{2}\hbar
N(0)v_{F}\Delta(0)$.} \label{fig4}
\end{figure}
\begin{figure}[tbp]
\includegraphics[width=1.1\columnwidth]{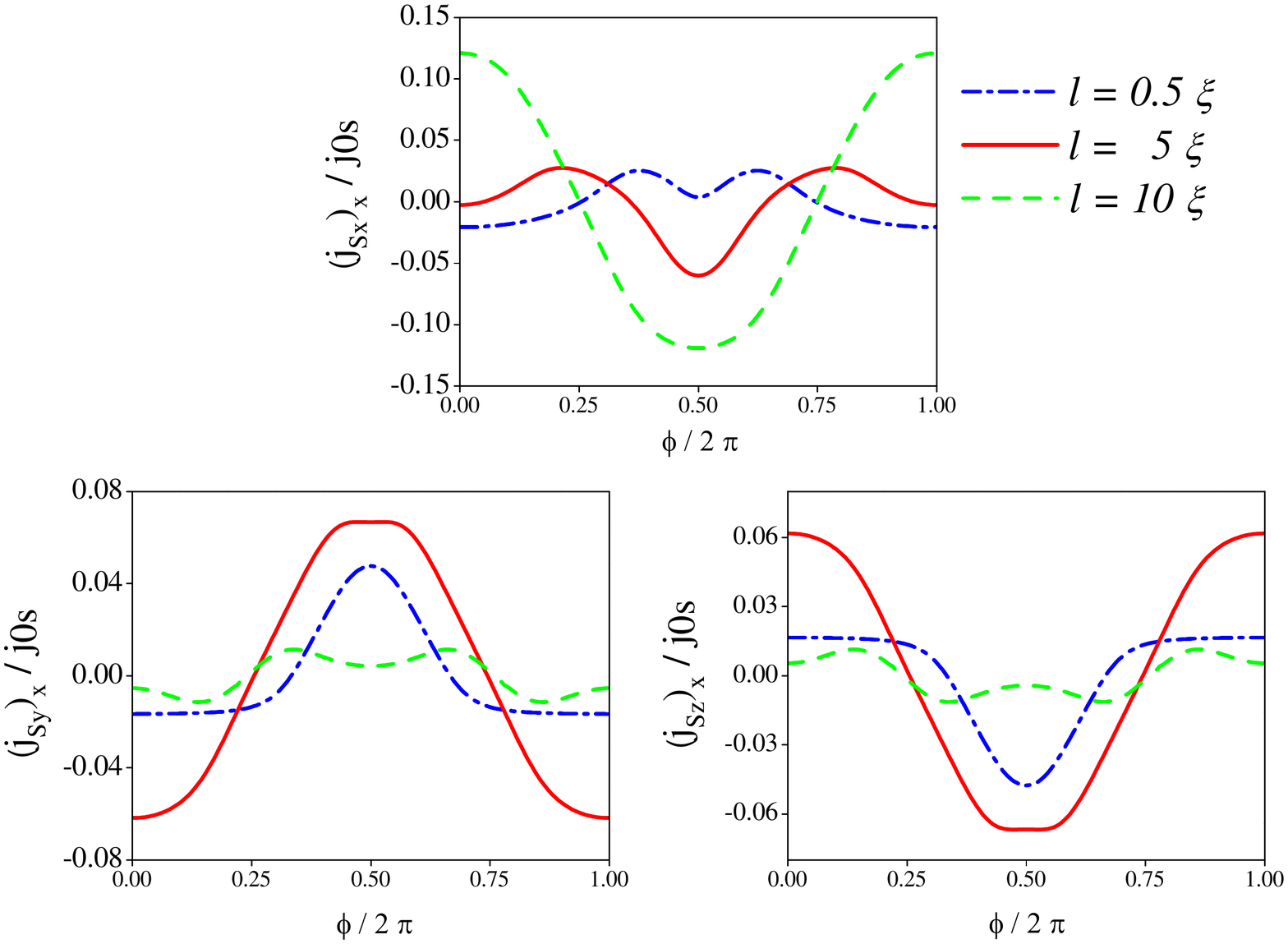}
\caption{(Color online) The $x$ component of the spin
current at the left interface, versus the phase difference $ \protect\phi $
for $h=5 \Delta$, $T=0.05T_{C}$, $\alpha=\pi/4$ and different $l$.}  \label{fig5}
\end{figure}

\section{analytical Results of Green functions}
\label{3}

 \newpage
 In principle, Eilenberger equations (\ref{Eilenberger}) should be supplemented with a self-consistent equation for gap vector. While these numerical self-consistent calculations can be used to investigate the spatial variation of gap
vector in the superconductors and also the dependence of
gap vector on temperature, in this paper we do not apply
them. In our analytical calculations, we consider a step-like
function of gap vector which is
zero in the ferromagnet and finite value in the superconducting part.
It has been shown that the absolute value of a self-consistently detemined order parameter decreases near the interface between
superconductor and normal metal, while its dependence on the
direction of Fermi velocity remains unaltered\cite{Barash}.
In Refs.\cite{Coury,Barash,Thuneberg,Viljas} a qualitative agreement
between self-consistent and non-self-consistent calculations for
unconventional Josephson junctions is found. Also, it is
observed that the results of non-self-consistent calculation in Ref.\cite{Faraii} are coincident with the experimental data in Ref.\cite{Freamat}. Also, the results by non-self-consistent treatment in Ref.\cite{Yip} are similar to the experimental results in Ref. \cite{Backhaus}.
Consequently, we believe that non-self-consistent approximation
doesn't change results qualitatively. Therefore, in our calculations, a
simple model of the constant order parameter up to the interface is
considered to obtain qualitative results. Using such a model, the analytical expressions for the charge and spin currents in the ferromagnet and superconductors can be obtained for a specified form of the order parameter. The
expressions for the charge and spin currents are the followings:\cite{Serene,Viljas}
\begin{equation}
\mathbf{j}_{e}\left( \mathbf{r}\right) =2i\pi eTN\left( 0\right)
\sum_{m}\left\langle \mathbf{v}_{F}g_{1}\left( \mathbf{\hat{v}}_{F},\mathbf{r%
},\varepsilon _{m}\right) \right\rangle  \label{charge-current},
\end{equation}
\begin{equation}
\mathbf{j}_{s_{i}}\left( \mathbf{r}\right) =\pi \hbar TN\left(
0\right) \sum_{m}\left\langle \mathbf{v}_{F}\left( \mathbf{{\hat{e}}}_{i}%
\cdot\mathbf{g}_{1}\left(
\mathbf{\hat{v}}_{F},\mathbf{r},\varepsilon _{m}\right) \right)
\right\rangle \label{spin-current}
\end{equation}
where $\mathbf{{\hat{e}}}_{i}\mathbf{=}\left( \hat{\mathbf{x}},\hat{%
\mathbf{y}},\hat{\mathbf{z}}\right) $ and $N(0)$ is the density of
states in the normal state at the Fermi energy.
 We consider the order parameter as follows:
\begin{equation}\label{axial}
\mathbf{d}_n(T,\mathbf{\hat{v}}_{F},\mathbf{r})=d
_{n}(T,\mathbf{\hat{v}}_{F},\mathbf{r})\hat{\mathbf{z}}.
\end{equation}
Solving Eilenberger equation in ferromagnet region and
superconductors under the continuity of solutions across the
interfaces, $x=0, l$ and the boundary conditions at the bulks, we
obtain the Green's functions in the ferromagnet region.
The Green's functions are given by
\begin{equation}\label{charge}
g_{1F}=\frac {\eta(
p^2-1)\left(p^2+2p[\beta+\sin^2\alpha(2\beta^2-\beta-1)]+1\right)}
{D},
\end{equation}
and
\begin{eqnarray}
 \mathbf{g}_{1F} =\frac{\eta p}
{D}\left(
\begin{array}{cc}
A\cos(\frac{2hx}{v_x})+B\sin(\frac{2hx}{v_x})\\
 \cos\alpha[ A\sin(\frac{2hx}{v_x})-B\cos(\frac{2hx}{v_x})]+C\sin\alpha\\
  \sin\alpha[ B\cos(\frac{2hx}{v_x})-A\sin(\frac{2hx}{v_x})]+C\cos\alpha
  \end{array}
  \right)\label{spin}
\end{eqnarray}
\begin{widetext}
where
\begin{eqnarray}\label{1}
\nonumber
A&=&\sin 2\alpha[\beta-1][p^2(2\beta+1)+2p(\beta\sin^2\alpha+\cos^2\alpha)-1],\\
\nonumber
B&=&\sin 2\alpha\sin(2hl/v_x)[p^2(2\beta-1)+2p(\beta\sin^2\alpha+\cos^2\alpha)+1],\\
\nonumber
C&=&2\cos^2\alpha\sin(2hl/v_x)[p^2+2p\left(\sin^2\alpha(\beta-1)+\beta\right)+1],\\
\nonumber
D&=&p^4+2p^2[2(\beta\sin^2\alpha+\cos^2\alpha)^2+2\beta^2-1]+4p\beta(p^2+1)(\beta\sin^2\alpha+\cos^2\alpha)+1,\\
\nonumber
\beta&=&\cos(2hl/v_x)\\
 \nonumber
\eta &=& sgn\left(v_{x}\right)\\
\nonumber p&=&\frac{d_1d_2^{*}\exp(-2\varepsilon_m
l/v_x)}{(\varepsilon_m+\eta\Omega_1)(\varepsilon_m+\eta\Omega_2)}\\
 \nonumber
\Omega_{1,2}&=&\sqrt{\varepsilon_{m}^{2}+\left|\mathbf{d}_{1,2}\right|^{2}}.
\end{eqnarray}
\end{widetext}
For parallel alignment of gap vector and exchange field, $\alpha=0$, Green's functions reduce to
\begin{eqnarray}\label{charge1}
\nonumber
g_{1F}&=&\frac {\eta( p^2-1)}  {p^2+2p \beta+1},\\
\mathbf{g}_{1F}&=&\frac{2\eta p\sin(2hl/v_x)}{p^2+2p
\beta+1}\mathbf{\hat{z}}.
\end{eqnarray}
It is seen from this expression that Green's functions and hence
charge and spin currents depend on the exchange field through
$\beta$.
Also, for the perpendicular configuration
$\alpha=\frac{\pi}{2}$, we obtain
\begin{eqnarray}\label{charge2}
\nonumber
g_{1F}&=&\frac {\eta( p-1)}{p+1},\\
\mathbf{g}_{1F}&=&0
\end{eqnarray}
where the Green's function are independent of the exchange field. As a result, we do not see any $0-\pi$ transition
in this case\cite{Powell}.

\begin{figure}[tbp]
\includegraphics[width=\columnwidth]{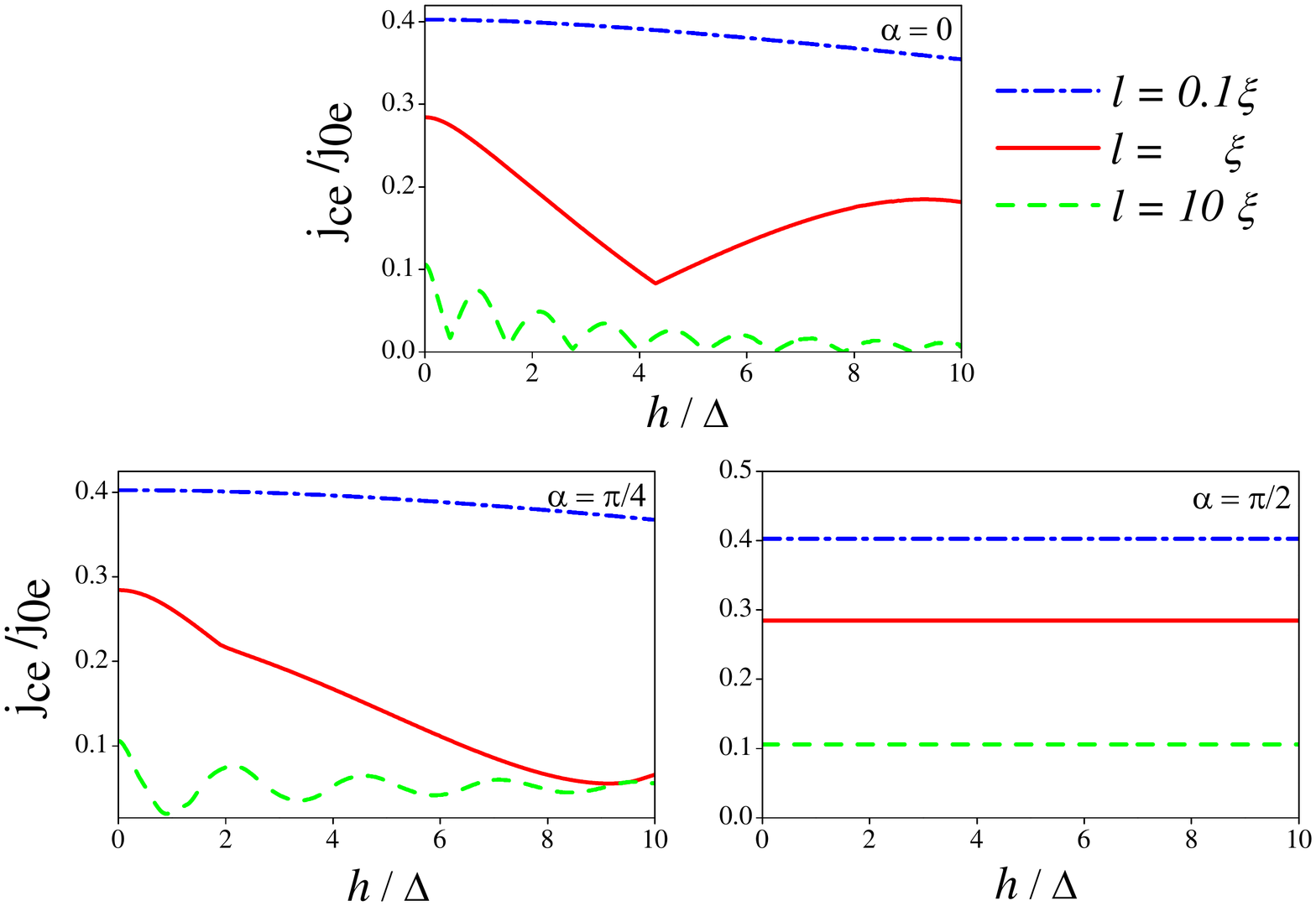}
\caption{(Color online) The $x$ component of the critical charge current
 versus $h$ for $T=0.05T_{C}$ and different $l$ and different $\alpha$ } \label{fig6}
\end{figure}

\begin{figure}[tbp]
\includegraphics[width=\columnwidth]{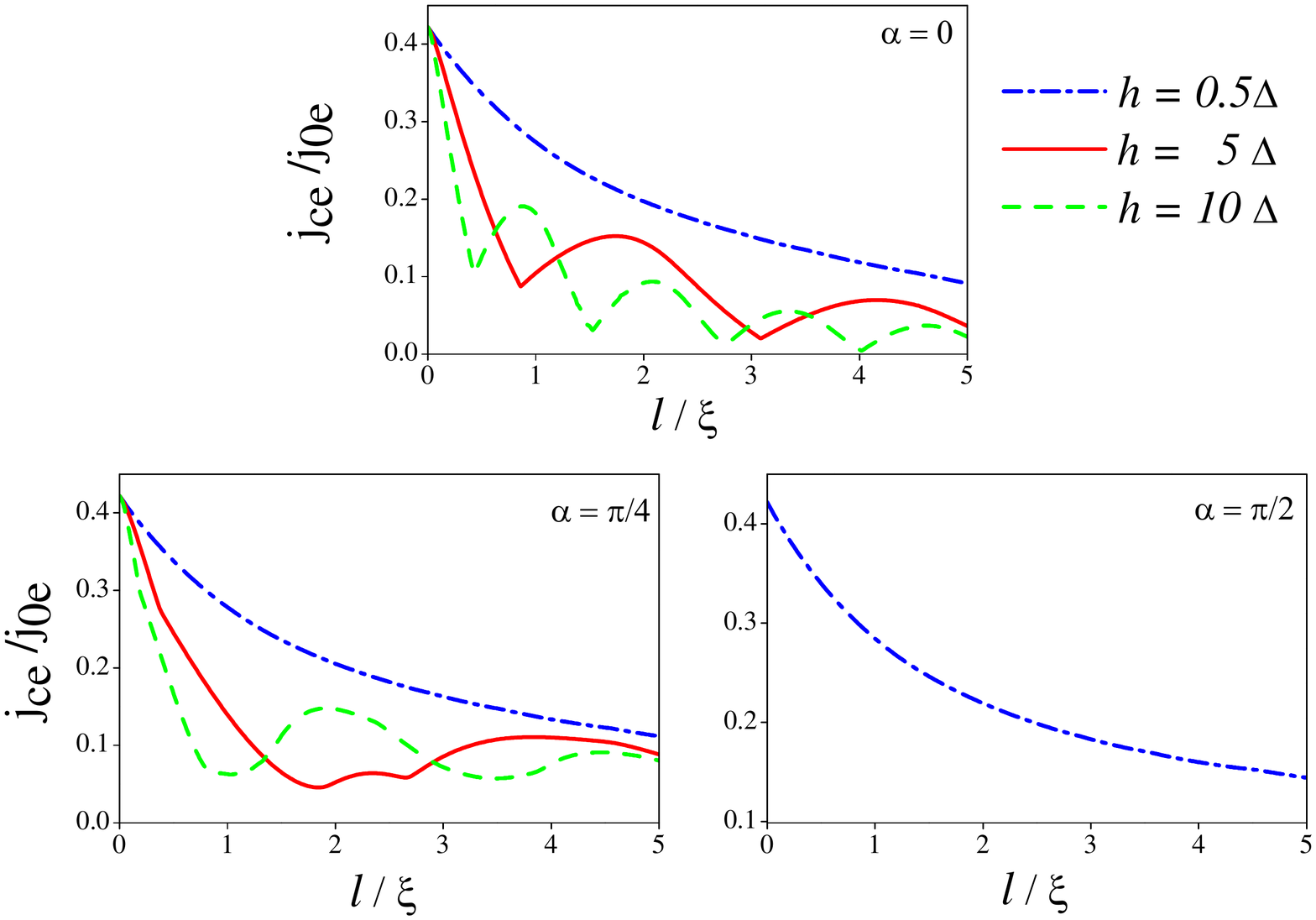}
\caption{(Color online) The $x$ component of the critical charge current
  versus the thickness of the ferromagnetic layer at $T=0.05T_{C}$
   for different $h$ and different $\alpha$} \label{fig7}
\end{figure}

\begin{figure}[tbp]
\includegraphics[width=1.2\columnwidth]{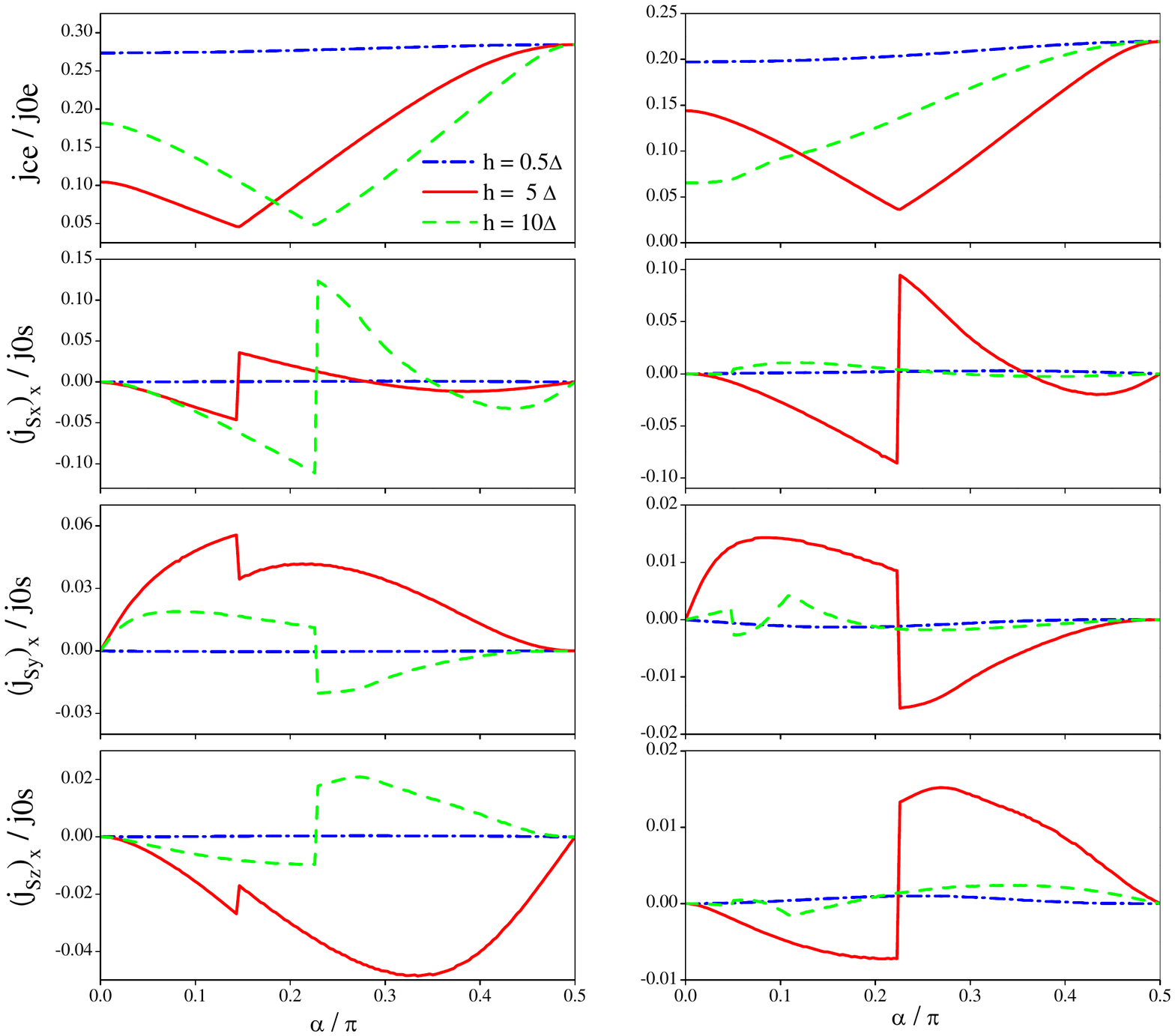}
\caption{(Color online) The $x$ component of the critical charge
and spin currents  versus $\alpha$ for $T=0.05T_{C}$.
Left panels are for $l=\xi$ and right panels are for $l=2 \xi$. } \label{fig8}
\end{figure}

\section{Numerical Results of Charge and Spin Currents}
\label{4} Here, we have considered $p-$wave superconductors with the
model of order parameter as
\begin{equation}
d(T,\mathbf{\hat{v}}_{F},\mathbf{r})=\Delta(T)\left(
k_{x}+ik_{y}\right)
\end{equation}
which would be realized in Sr$_{2}$RuO$_{4}$\cite{Mackenzie}. The
function $\Delta\left( T\right) $ describes the dependence of the
order parameter $\mathbf{d}$ on the temperature $T$. We have used
Green functions in Eqs.(\ref{charge}) and (\ref{spin}) and
calculated charge and spin currents for our specified model of the
order parameter. For representative values of $h$, $l$ and $\alpha$,
we have shown charge and spin currents in Figs.
\ref{fig2}-\ref{fig5} as a function of phase difference between
superconductors. In the figures, $\xi(T)=\frac{\hbar v_F}{\pi
\Delta(T)}$ denotes a superconductor coherence length at the bulk of
superconductor.
 By changing the parameters, we find a tendency towards the 0-$\pi$ transition.
 In particular, we see a stronger tendency towards the 0-$\pi$ transition  at $\alpha=0$.

In our formalism, there are in general three components ($s_x$, $s_y$ and $s_z$) of spin current due to the presence of the ferromagnetic layer with finite thickness.
While charge current is independent of position in ferromagnet, spin
current is dependent on position and is not conserved. In fact, spin current at the left and right interfaces have opposite sign and in the middle of ferromagnetic layer, the spin current vanishes.
We have plotted spin current at the left interface between superconductor
and ferromagnetic layer in Figs. \ref{fig4},\ref{fig5} and
Fig.\ref{fig8}. At this point, there are three components of spin
current.


For charge current, we see that by increasing the  thickness of
ferromagnetic layer the current decreases with oscillation because of
decreasing of quantum coherency, but for the spin current the
situation is quite different. As a result of the exchange field, the
thicker layer has a greater effect on spin of quasiparticles and
thus a larger spin current. As seen in Fig.\ref{fig2} for
$\alpha=\pi/2$, the results are independent of the exchange field.
Thus, charge current in this case is only dependent on thickness of
the ferromagnetic layer and spin current is absent. Regarding spin
currents, all components of spin current vanish at $\alpha=0$ and
$\alpha=\pi/2$.
As shown in Fig.\ref{fig4}, spin current is strongly dependent on the
exchange field, so for a weak ferromagnet (for example
$h=0.5\Delta$), spin current is negligibly small.
For some values of the phase difference, spin current exists in the absence
of charge current as seen from Figs. \ref{fig4} and \ref{fig5}. Also, there is local maximum or minimum in spin
current in $\phi=0$ or $\phi=\pi$. This can be explained as follows.
Charge currents are generally odd function but spin currents are
even function of the phase difference, because of the symmetry under
the time reversal. Therefore, the slope of the spin current becomes
zero at $\phi=0$ and $\phi=\pi$.

The critical currents as a function of $h$ and $l$ are shown in
Fig.\ref{fig6} and Fig.\ref{fig7} for different misorientations
$\alpha=0$, $\pi/4$ and $\pi/2$. Also, critical charge and spin
currents as the function of $\alpha$ have been plotted in
Fig.\ref{fig8}. Here, the critical charge and spin currents are
defined as $j_{ce}=Max_{\phi}j_{e}(\phi)=j_{e}(\phi^{\ast})$ and
$j_{s}=j_{s}(\phi^{\ast})$, respectively. The 0-$\pi$ transition as
a function of $h$ for large $l$ and small $\alpha$ are shown in
Fig.\ref{fig6}. In Ref.\cite{Brydon2}, it is shown that 0-$\pi$
transition cannot occur at $\alpha=\pi/2$ for a $\delta$ function
type ferromagnetic barrier. We find that this is the case even in
the presence of a ferromagnetic layer with finite thickness instead of
a $\delta$ function type barrier.
This is also seen from the dependence of the Josephson current on
$l$ which shows the 0-$\pi$ transition for large $h$ and small
$\alpha$ as seen in Fig.\ref{fig7}. As shown in Fig.\ref{fig8}, for
finite thickness of ferromagnetic layer, 0-$\pi$ transition can be
found with changing the misorinetation angle. The 0-$\pi$ transition
is accompanied by the jump of the spin current due to the phase jump
at the 0-$\pi$ transition.\cite{Alidoust} Even when charge current
shows a monotonous dependence on $\alpha$, spin current can show
non-monotonous $\alpha$ dependence.

\section{Conclusions}\label{5}
In this paper, we have investigated transport properties in triplet SFS Josephson junction
and clarified the effects of the thickness of ferromagnet, the exchange field, and misorientation
angle between exchange field of ferromagnet and $d$-vector of superconductors on the Josephson
charge and spin currents. We found that 0-$\pi$ transition can occur except for the case that the exchange field and d-vector are in nearly perpendicular configuration.
Also, we showed how spin current flows due to misorientation between the exchange field and
$d$-vector and found that it is absent when two vectors are parallel or perpendicular to each other.

\section{Acknowledgement}\nonumber
We would like to thank J. Linder and P. M. R. Brydon for their helpful discussions and their comments on draft of our paper.


\begin{thebibliography}{99}
\bibitem{buzdinrmp} A. I. Buzdin, Rev. Mod. Phys. \textbf{77}, 935 (2005).

\bibitem{Bergeret1} F. S. Bergeret, A. F. Volkov, and K. B. Efetov, Rev. Mod. Phys. \textbf{77}, 1321 (2005).

\bibitem{Volkov} A. F. Volkov, and K. B. Efetov, Phys. Rev. B. \textbf{78}, 024519 (2008).

\bibitem{Zareyan1} M. Zareyan, W. Belzig, and Yu. V. Nazarov, Phys. Rev. B. \textbf{65}, 184505 (2002).

\bibitem{Braude} V. Braude and Ya. M. Blanter, Phys. Rev. Lett. \textbf{100}, 207001 (2008).

\bibitem{Efetov2} F. S. Bergeret, A. F. Volkov, and K. B. Efetov, Phys. Rev.
Lett. \textbf{86}, 4096 (2001); A. F. Volkov, F. S. Bergeret, and K. B. Efetov, Phys. Rev. Lett. \textbf{90}, 117006 (2003).

\bibitem{Kadigro} A. Kadigrobov, R. I. Shekter and M. Jonson, Europhys.
Lett. \textbf{90}, 394 (2001).

\bibitem{EschrigLTP} M. Eschrig, T. L\"ofwander, Th. Champel, J. C. Cuevas, and G. Sc\"hon, J. Low Temp. Phys. \textbf{147} 457 (2007).

\bibitem{eschrig} M. Eschrig and T. L\"ofwander, Nature Physics \textbf{4}, 138 (2008).

\bibitem{Yokoyama2}T. Yokoyama, Y. Tanaka, and A. A. Golubov, Phys. Rev. B \textbf{75}, 134510 (2007);T. Yokoyama and Y. Tserkovnyak, Phys. Rev. B \textbf{80}, 104416 (2009).

\bibitem{Linder} J. Linder, T. Yokoyama, A. Sudb{\o}, and M. Eschrig, Phys. Rev. Lett. \textbf{102}, 107008 (2009); J. Linder, T. Yokoyama, and A. Sudb{\o}, Phys. Rev. B \textbf{79}, 054523 (2009);J. Linder, A. Sudbo, T. Yokoyama, R. Grein and M. Eschrig, Phys. Rev. B \textbf{81}, 214504 (2010).


\bibitem{Alidoust} M. Alidoust, J. Linder, G. Rashedi, T. Yokoyama, and A. Sudbo Phys. Rev. B. \textbf{81}, 014512 (2010).

\bibitem{Maeno} Y. Maeno, H. Hashimoto, K. Yoshida, S. Nishizaki, T. Fujita, J. G. Bednorz, and F. Lichtenberg, Nature (London) \textbf{372}, 532 (1994).


\bibitem{Ishida} K. Ishida, H. Mukuda, Y. Kitaoka, K. Asayama, Z. Q. Mao, Y. Mori, and Y. Maeno, Nature (London) \textbf{396}, 658 (1998).

\bibitem{Luke} G. M. Luke, Y. Fudamoto, K. M. Kojima, M. I. Larkin, J. Merrin, B. Nachumi, Y. J. Uemura, Y. Maeno, Z. Q. Mao, Y. Mori, H. Nakamura, and M. Sigrist, Nature (London) \textbf{394}, 558 (1998).


\bibitem{Mackenzie} A. P. Mackenzie and Y. Maeno, Rev. Mod. Phys. \textbf{75}, 657 (2003).
\bibitem{Nelson} K. D. Nelson, Z. Q. Mao, Y. Maeno and Y. Liu, Science \textbf{306}, 1151 (2004).
\bibitem{Asano1} Y. Asano, Y. Tanaka, M. Sigrist and S. Kashiwaya, Phys. Rev. B \textbf{67}, 184505 (2003);
 Phys. Rev. B \textbf{71}, 214501 (2005).
\bibitem{Tou} H. Tou, Y. Kitaoka, K. Ishida, K. Asayama, N. Kimura, Y. Onuki, E. Yamamoto, Y. Haga and
K. Maezawa, Phys. Rev. Lett. \textbf{80}, 3129 (1998).
\bibitem{Muller} V. Muller, Ch. Roth, D. Maurer, E. W. Scheidt, K. Lers, E. Bucher and H. E. Bmel, Phys. Rev. Lett. \textbf{58}, 1224 (1987).
\bibitem{Qian} Y. J. Qian, M. F. Xu, A. Schenstrom, H. P. Baum, J. B. Ketterson, D. Hinks,
M. Levy and B. K. Sarma, Solid State Commun., \textbf{63}, 599
(1987).
\bibitem{Abrikosov} A. A. Abrikosov, J. of Low Temp. Phys.  \textbf{53}, 359 (1983).
\bibitem{Fukuyama} H. Fukuyama and Y. Hasegawa, J. Phys. Soc. Jpn \textbf{56}, 877 (1987).
\bibitem{Lebed} A. G. Lebed, K. Machida and M. Ozaki, Phys. Rev. B \textbf{62}, 795 (2000).

\bibitem{Saxena} S. S. Saxena, P. Agarwal, K. Ahilan, F. M. Grosche, R. K. W. Haselwimmer, M. J. Steiner, E. Pugh, I. R. Walker, S. R. Julian, P. Monthoux, G. G. Lonzarich, A. Huxley, I. Shelkin, D. Braithwaite, and J. Flouquet, Nature (London) \textbf{406}, 587 (2000).

\bibitem{Pfleiderer} C. Pfleiderer, M. Uhlarz, S. M. Hayden, R. Vollmer, H. v. Lohneysen, N. R. Bernhoeft, and G. G. Lonzarich, Nature (London) \textbf{412}, 58 (2001).

\bibitem{Aoki} D. Aoki, A. Huxley, E. Ressouche, D. Braithwaite, J. Flouquet, J. Brison, E. Lhotel, and C. Paulsen, Nature (London) \textbf{413}, 613 (2001).

\bibitem{Bolech} C. J. Bolech and T. Giamarchi, Phys. Rev. Lett. \textbf{92}, 127001 (2004); Phys. Rev. B \textbf{71}, 024517 (2005).

\bibitem{Gronsleth} M. S. Gr{\o}nsleth, J. Linder, J.-M. B{\o}rven, and A. Sudb{\o}, Phys. Rev. Lett. \textbf{97},
147002 (2006); Phys. Rev. B \textbf{75}, 054518 (2007).

\bibitem{Yasuhiro} Y. Asano, Phys. Rev. B \textbf{72}, 092508 (2005); Phys. Rev. B \textbf{74}, 220501(R) (2006).
\bibitem{Mahmoodi} R. Mahmoodi, Yu. A. Kolesnichenko and S. N. Shevchenko, Fiz. Nizk. Temp. \textbf{28} 262 (2002) [Low Temp. Phys. \textbf{28} 184 (2002)].
\bibitem{Kolesnichenko1} G. Rashedi  and Yu A. Kolesnichenko, Physica C, \textbf{451}
31 (2007); G. Rashedi  and Yu A. Kolesnichenko, Supercond. Sci.
Technol. \textbf{18}, 482 (2005).

\bibitem{Rashedi1} G. Rashedi, Y. Rahnavard, Yu. A. Kolesnichenko, Fiz. Nizk. Temp. \textbf{36}
262 (2010) [Low Temp. Phys. \textbf{36}, 205 (2010)].
\bibitem{Rashedi2} G. Rashedi and
Yu. A. Kolesnichenko, Fiz. Nizk. Temp. \textbf{31} 634 (2005) [Low
Temp. Phys. \textbf{31}, 481 (2005)]

\bibitem{Rahnavard} Y. Rahnavard, G. Rashedi and T. Yokoyama, J. Phys. Condens. Matter \textbf{22}, 415701(2010) .

\bibitem{Kastening} B. Kastening, D. K. Morr, D. Manske, and K. Bennemann, Phys.
Rev. Lett. \textbf{96}, 047009 (2006).
\bibitem{Brydon1} P. M. R. Brydon, B. Kastening, D. K. Morr, and D. Manske,
Phys. Rev. B. \textbf{77}, 104504 (2008).

\bibitem{Brydon2} P. M. R. Brydon and D. Manske, Phys. Rev. Lett. \textbf{103}, 147001 (2009); P. M. R. Brydon, C Iniotakis, and Dirk Manske, New Journal of Physics \textbf{11}, 055055 (2009).

\bibitem{Eilenberger} G. Eilenberger, Z. Phys. \textbf{214}, 195 (1968).

\bibitem{Kulik} I. O. Kulik, and A. N. Omelyanchuk,, Fiz. Nizk. Temp. \textbf{4}, 296 (1978)[Sov. J. Low Temp. Phys. \textbf{4}, 142 (1978)].

\bibitem{Kupriyanov}  M. Yu. Kupriyanov, Fiz. Nizk. Temp. \textbf{7}, 700 (1981)[Sov. J. Low. Temp. Phys. \textbf{7},
342 (1981)].
\bibitem{Barash} Yu. S. Barash, A.M. Bobkov, and M. Fogelstr\"{o}m, Phys. Rev. B \textbf{64}, 214503 (2001).
\bibitem{Coury} M.H.S. Amin, M.Coury, S.N. Rashkeev, A.N. Omelyanchouk, and A.M. Zagoskin, Physica B,
\textbf{318}, 162 (2002).
\bibitem{Thuneberg} J.K. Viljas, E.V. Thuneberg, Phys. Rev. B \textbf{65}, 64530 (2002).
\bibitem{Viljas} J. Viljas, cond-mat/0004246.
\bibitem{Faraii} Z. Faraii and M. Zareyan, Phys. Rev. B \textbf{69}, 014508(2004).
\bibitem{Freamat} M. Freamat, K.-W. Ng, Phys. Rev. B \textbf{68}, 060507 (2003).
\bibitem{Yip} S.-K. Yip, Phys. Rev. Lett. \textbf{83}, 3864 (1999).
\bibitem{Backhaus} S. Backhaus, S. Pereverzev, R.W. Simmonds, A. Loshak, J.C. Davis, and
R.E. Packard, Nature \textbf{392}, 687 (1998).
\bibitem{Serene} J. Serene and  D. Rainer, Phys. Reports \textbf{101}, 221 (1983).
\bibitem{Powell} B. J. Powell, J. F. Annett and B. L. Gy\"{o}rffy, J. Phys. A: Math. Gen. \textbf{36}
9289 (2003).

\end{thebibliography}
\end{document}